\documentclass[11pt]{article}
\usepackage{latexsym}
\usepackage{graphicx}% Include figure files
\def\ltsim{\lower3pt\hbox{$\, 
\buildrel < \over \sim \, $}}  
\def\gtsim{\lower3pt\hbox{$\, \buildrel > \over \sim \, $}}  
\makeatletter
\def\section{\@startsection {section}{1}{\z@}{-3.5ex plus -1ex minus
 -.2ex}{2.3ex plus .2ex}{\large\bf}}
\def\subsection{\@startsection{subsection}{2}{\z@}{-3.25ex plus -1ex
minus -.2ex}{1.5ex plus .2ex}{\normalsize\bf}}
\makeatother
\makeatletter
\def\theequation{\arabic{section}.\arabic{equation}}

\@addtoreset{equation}{section}
\renewcommand{\theequation}{\thesection.\arabic{equation}}
\makeatother
\newcommand{\captionfonts}{\small}
\makeatletter  % Allow the use of @ in command names
\long\def\@makecaption#1#2{%
  \vskip\abovecaptionskip
  \sbox\@tempboxa{{\captionfonts #1: #2}}%
  \ifdim \wd\@tempboxa >\hsize
    {\captionfonts #1: #2\par}
  \else
    \hbox to\hsize{\hfil\box\@tempboxa\hfil}%
  \fi
  \vskip\belowcaptionskip}
\makeatother   % Cancel the effect of \makeatletter
\catcode`@=11
\def\marginnote#1{}
\newcount\hour
\newcount\minute
\newtoks\amorpm
\hour=\time\divide\hour
by60
\minute=\time{\multiply\hour by60 \global\advance\minute
by-\hour}
\edef\standardtime{{\ifnum\hour<12 \global\amorpm={am}
\else\global\amorpm={pm}\advance\hour by-12 \fi
 \ifnum\hour=0
\hour=12 \fi
 \number\hour:\ifnum\minute<10
0\fi\number\minute\the\amorpm}}
\edef\militarytime{\number\hour:\ifnum\minute<10
0\fi\number\minute}
\def\draftlabel#1{{\@bsphack\if@filesw
{\let\thepage\relax
 \xdef\@gtempa{\write\@auxout{\string
\newlabel{#1}{{\@currentlabel}{\thepage}}}}}\@gtempa
 \if@nobreak
\ifvmode\nobreak\fi\fi\fi\@esphack}
\gdef\@eqnlabel{#1}}
\def\@eqnlabel{}
\def\@vacuum{}
\def\draftmarginnote#1{\marginpar{\raggedright\scriptsize\tt#1}}
\def\draft{\oddsidemargin
0.0truein
 \def\@oddfoot{\sl preliminary draft \hfil
\rm\thepage\hfil\sl\today\quad\militarytime}
 \let\@evenfoot\@oddfoot
\overfullrule 3pt
 \let\label=\draftlabel
\let\marginnote=\draftmarginnote
\def\@eqnnum{(\theequation)\rlap{\kern\marginparsep\tt\@eqnlabel}
\global\let\@eqnlabel\@vacuum}
}
\catcode`@=12
\def\XXint#1#2#3{{\setbox0=\hbox{$#1{#2#3}{\int}$}
     \vcenter{\hbox{$#2#3$}}\kern-.5\wd0}}

\def\bea{\begin{eqnarray}} \def\eea{\end{eqnarray}}
\def\be{\begin{eqnarray}} \def\ee{\end{eqnarray}} \def\nn{\nonumber}
\newcommand{\promille}{%
  \relax\ifmmode\promillezeichen
        \else\leavevmode\(\mathsurround=0pt\promillezeichen\)\fi}
\newcommand{\promillezeichen}{%
  \kern-.05em%
  \raise.5ex\hbox{\the\scriptfont0 0}%
  \kern-.15em/\kern-.15em%
  \lower.25ex\hbox{\the\scriptfont0 00}}
\begin{document}

\thispagestyle{empty}

\begin{center}
\hfill IFT-UAM/CSIC-08-09\\
\hfill CERN-PH-TH/2008-033\\
\hfill UAB-FT-639

\begin{center}

\vspace{1.7cm}

{\LARGE\bf The Higgs as a Portal to\\\vspace*{0.5cm}Plasmon-like 
Unparticle 
Excitations}

\end{center}

\vspace{1.4cm}

{\bf A. Delgado$^{\,a}$, J. R. Espinosa$^{\,b,c}$, J. M. 
No$^{\,b}$ and M. Quir\'os$^{\,d}$}\\

\vspace{1.2cm}

${}^a\!\!$
{\em {Department of Physics, 225 Nieuwland Science Hall, U. of Notre 
Dame,\\ Notre Dame, IN 46556-5670, USA}}

${}^b\!\!$
{\em { IFT-UAM/CSIC, Fac. Ciencias UAM, 28049 Madrid, Spain}}

${}^c\!\!$
{\em { CERN, Theory Division, CH-1211, Geneva 23, Switzerland}}

${}^d\!\!$
{\em { Instituci\`o Catalana de Recerca i Estudis Avancats (ICREA)} at}

{\em {IFAE, Universitat Aut{\`o}noma de Barcelona,
08193 Bellaterra, Barcelona, Spain
}}

\end{center}

\vspace{0.8cm}

\centerline{\bf Abstract}
\vspace{2 mm}
\begin{quote}\small

A renormalizable coupling between the Higgs and a scalar unparticle
operator $\mathcal O_U$ of non-integer dimension $d_U<2$ triggers,
after electroweak symmetry breaking, an infrared divergent vacuum
expectation value for $\mathcal O_U$. Such IR divergence should be
tamed before any phenomenological implications of the Higgs-unparticle
interplay can be drawn. In this paper we present a novel mechanism to
cure that IR divergence through (scale-invariant) unparticle
self-interactions, which has properties qualitatively different from
the mechanism considered previously. Besides finding a mass gap in the
unparticle continuum we also find an unparticle pole reminiscent of a
plasmon resonance. Such unparticle features could be explored 
experimentally
through their mixing with the Higgs boson.
\end{quote}

\vfill

\newpage
\section{Introduction}

The very active field of unparticles grew out of two seminal
papers~\cite{Georgi} in which Georgi entertained the possibility of
coupling a scale-invariant sector (with a non-trivial infrared fixed
point) to our familiar standard model of particles.  He described
several very unconventional features of that sector that could be
probed through such couplings. In his original proposal, Georgi
considered only couplings through non-renormalizable operators (after
integrating out some heavy messenger sector that interacts directly
both with the Standard Model and the unparticle sector).
Later on Shirman et al.~\cite{shirman} considered the possibility of
coupling directly a scalar operator of unparticles ${\cal O}_U$ (of
scaling dimension $d_U$, with $1<d_U<2$) to the SM Higgs field through
a renormalizable coupling of ${\cal O}_U$ to $|H|^2$. As pointed out
in~\cite{DEQ} such coupling induces a tadpole for ${\cal O}_U$ after
the breaking of the electroweak symmetry and for $d_U<2$ the value of
the vacuum expectation value $\langle {\cal O}_U\rangle$ has an
infrared (IR) divergence. That divergence should be cured before any
phenomenological implications of the Higgs-unparticle coupling can be
studied in a consistent way. Ref.~\cite{DEQ} discussed a simple way of
inducing an IR cutoff that would make $\langle {\cal O}_U\rangle$
finite. One of the main implications of such mechanism was the
appearance of a mass gap, $m_g$, of electroweak size for the
unparticle sector. Needless to say, such mass gap has dramatic
implications both for phenomenology and for constraints on the
unparticle sector.

In addition, Ref.~\cite{DEQ} showed that, after electroweak symmetry
breaking (EWSB), the Higgs state mixes with the unparticle continuum
in a way reminiscent of the Fano-Anderson model~\cite{FA}, familiar in
solid-state and atomic physics as a description of the mixing between
a localized state and a quasi-continuum. When the Higgs mass is below
$m_g$, the Higgs survives as an isolated state but with some
unparticle admixture that will modify its properties. On the other
hand, the unparticle continuum above $m_g$ gets a Higgs contamination
that can make it more accessible experimentally. When the Higgs mass is
above $m_g$ the Higgs state gets subsumed into the unparticle
continuum with the Higgs width greatly enlarged by the unparticle
mixing. Such behaviour is similar to that found when the Higgs mixes
with a quasi-continuum of graviscalars~\cite{gravis}.  In both cases,
with $m_h$ above or below $m_g$, the properties of the mixed
Higgs-unparticle system can be described quite neatly through a
spectral function analysis.

The organization of the paper is as follows: after describing the
previous IR problem (Section~2) we present an alternative
stabilization mechanism for $\langle {\cal O}_U\rangle$
(Section~3). This mechanism has significant differences with respect
to that used in Ref.~\cite{DEQ}: although it also induces an
unparticle mass gap\footnote{One expects such mass gap as a generic
feature of any mechanism that solves the IR problem.} it involves a
scale-invariant self-coupling of unparticles only and leads to the
appearance of a peculiar resonance in the unparticle
continuum that is reminiscent of a plasmon excitation
(Section~4). The mixing between the unparticle states and the Higgs
boson after EWSB gives a handle on the structure of the
unparticle continuum. This is best seen in terms of an spectral
function analysis which we develop in Section~5. We present our
conclusions in Section~6.

\section{The Infrared Problem}

We start with the following scalar potential
\be
V_0=m^2 |H|^2+\lambda|H|^4+\kappa_U
|H|^2\mathcal O_{U}\ ,
\label{tree}
\ee
where the first two terms are the usual SM Higgs potential and the
last term is the Higgs-unparticle coupling, with $\kappa_U$ having
mass dimension $2-d_U$. As usual, the quartic coupling $\lambda$ would
be related in the SM to the Higgs mass at tree level by
$m_{h0}^2=2\lambda v^2$. We write the Higgs real direction as
$Re(H^0)=(h^0+v)/\sqrt{2}$, with $v =246$ GeV.

The unparticle operator $\mathcal O_U$ has dimension $d_U$, spin zero and 
its propagator is~\cite{Georgi,Cheung}
\be
P_U(p^2)=\frac{A_{d_U}}{2\sin(\pi d_U)} 
\frac{i}{(-p^2-i\epsilon)^{2-d_U}},\quad
A_{d_U}\equiv
\frac{16\pi^{5/2}}{(2\pi)^{2d_U}}\frac{\Gamma(d_U+1/2)}
{\Gamma(d_U-1)\Gamma(2d_U)}\ .
\label{prop}
\ee

When the Higgs field gets a non zero vacuum expectation value (VEV)
the scale invariance of the unparticle sector is
broken~\cite{shirman}.  From (\ref{tree}) we see that in such non-zero
Higgs background the physical Higgs field mixes with the unparticle
operator ${\cal O}_U$ and also a tadpole appears for ${\cal O}_U$
itself which will therefore develop a non-zero VEV.

As done in Ref.~\cite{DEQ}, it is very convenient to use a
deconstructed version of the unparticle sector as proposed
in~\cite{deco}. One considers an infinite tower of scalars $\varphi_n$
($n=1,...,\infty$) with masses squared $M_n^2=\Delta^2 n$. The mass
parameter $\Delta$ is small and eventually taken to zero, limit in
which one recovers a (scale-invariant) continuous mass spectrum. As
explained in~\cite{deco}, the deconstructed form of the operator
${\cal O}_U$ is
\be
\label{OUdec}
{\cal O}\equiv \sum_n F_n \varphi_n \ ,
\ee
where $F_n$ is chosen as
\be
\label{Fdec}
F_n^2 = \frac{A_{d_U}}{2\pi}\Delta^2 (M_n^2)^{d_U-2}\ , 
\ee 
so that the two-point correlator of ${\cal O}$ matches that of ${\cal
O}_U$ in the $\Delta\to 0$ limit. In the deconstructed theory the
unparticle scalar potential, including the coupling (\ref{tree}) to
the Higgs field, reads
\be
\label{potdec}
\delta{V} = \frac{1}{2}\sum_n M_n^2\varphi_n^2+\kappa_U |H|^2\sum_n F_n 
\varphi_n\ .
\ee
A non-zero VEV, $\langle |H|^2\rangle=v^2/2$, triggers a VEV for the
fields $\varphi_n$:
\be
\label{vn}
v_n\equiv\langle\varphi_n\rangle=-\frac{\kappa_U v^2}{2M_n^2}F_n\ ,
\ee
thus implying, in the continuum limit, 
\be
\label{vevOU}
\langle {\cal O}_U \rangle = -\frac{\kappa_U
v^2}{2} \int_0^\infty \frac{F^2(M^2)}{M^2}dM^2\ ,
\ee
where 
\be
\label{F}
F^2(M^2)=\frac{A_{d_U}}{2\pi}(M^2)^{d_U-2}\ ,
\ee
is the continuum version of (\ref{Fdec}). We see that
$\langle {\cal O}_U \rangle$ has an IR divergence for $d_U<2$, due to the
fact that for $M\to 0$ the tadpole diverges while the mass itself,
that should stabilize the unparticle VEV, goes to zero.

In Ref.~\cite{DEQ} it was shown how one can easily get an IR regulator in 
(\ref{F}) by including a coupling
\be
\label{HHUU}
\delta V = \zeta |H|^2 \sum_n \varphi_n^2\ ,
\ee
in the deconstructed theory. This coupling respects the conformal
symmetry but will break it when $H$ takes a VEV. Now one gets
\be
\langle {\cal O}_U \rangle = -\frac{\kappa_U v^2}{2} \int_0^\infty
\frac{F^2(M^2)}{M^2+\zeta v^2}\ dM^2\ ,
\ee
which is obviously finite for $1<d_U<2$ and reads explicitly
\be
\langle {\cal O}_U \rangle
=-\frac{1}{2}\kappa_U\frac{A_{d_U}}{2\pi}\zeta^{d_U-2}
v^{2d_U-2}\Gamma(d_U-1)\Gamma(2-d_U)\ .
\label{vevOUIR}
\ee
Implications for EWSB of such coupling (\ref{HHUU}) were 
studied in Ref.~\cite{DEQ}. 

\section{An Alternative Solution to the IR Problem}

It is natural to attempt to solve the IR problem of the previous
section by introducing a quartic coupling term for the
deconstructed scalar fields $\varphi_n$ so that the VEVs $v_n$ are
under control. As pointed out already in Ref.~\cite{DEQ} the naive try
with $\delta V=\lambda_U\sum_n \varphi_n^4$ fails. Here we prove that
the particular combination
\be \delta V = \frac{1}{4} \xi
\left(\sum_{n=1}^{\infty}\varphi_n^2\right)^2\ ,
\label{quartic}
\ee
is successful in providing a finite value for $\langle {\cal
O}_U\rangle$. Before showing that explicitly, let us first show that
the coupling (\ref{quartic}) has a finite and scale-invariant
continuum limit.

We can take as scale transformations for the deconstructed fields 
$\varphi_n$ 
\be
\varphi_n(x) \rightarrow a\varphi_n(x a)\ ,
\label{scale}
\ee
while leaving the space-time coordinates unscaled ($x\rightarrow x$). 
It is straightforward to show that under such scale transformation the
kinetic part of the (deconstructed) action is invariant while the mass
terms are not, as usual. In the continuum limit, however, taking 
$\Delta\cdot
u(M^2,x)$ as the continuum limit of $\varphi_n(x)$, and using the
scale transformation
\be u(M^2,x) \rightarrow u(M^2/a^2, x a)\ ,
\label{scalec}
\ee
the continuum action
\be
S=\int d^4 x \int_0^\infty dM^2 \left[\frac{1}{2}\partial_\mu 
u(M^2,x)\partial^\mu u(M^2,x)-M^2 u^2(M^2,x)\right]\ ,
\label{actionc}
\ee
is indeed scale invariant. Using the same construction, it is then
straightforward to see that the continuum limit of the quartic
coupling (\ref{quartic}) is well defined and scale invariant, being
explicitly given by: 
\be \delta S = -\int d^4 x \int_0^\infty dM_1^2
\int_0^\infty dM_2^2\ \frac{1}{4}\xi\ u^2(M_1^2,x) u^2(M_2^2,x)\ .
\ee

To keep the following analysis general, we consider both couplings
$\zeta$ and $\xi$ simultaneously, writing for the deconstructed part
of the scalar potential:
\be
\delta{V} = \frac{1}{2}\sum_n M_n^2\varphi_n^2+\kappa_U |H|^2\sum_n F_n 
\varphi_n + \zeta |H|^2 \sum_n \varphi_n^2 + \frac{1}{4} \xi 
\left(\sum_{n=1}^{\infty}\varphi_n^2\right)^2\ .
\label{Vfull}
\ee
The minimization equation for the Higgs field is not affected by the
new coupling $\xi$, while that for $v_n\equiv\langle\varphi_n\rangle$
can be put in the form
\be
v_n=\frac{-\frac{1}{2}\kappa_U v^2 F_n}{M_n^2+\zeta 
v^2+\xi\sum_{m=1}^\infty v_m^2}\ .
\ee
Squaring the above equation and summing in $n$ from 1 to $\infty$ one
gets an implicit equation for
\be
\sigma^2\equiv \sum_{n=1}^\infty v_n^2\ .
\label{sigmad}
\ee
In the continuum limit, and using 
\be
(\mu_U^2)^{2-d_U}\equiv \kappa_U^2 \frac{A_{d_U}}{2\pi}\ ,
\ee
the equation for $\sigma^2$ reads 
\be
 \sigma^2 = \frac{1}{4} (\mu_U^2)^{2-d_U} v^4 \int_0^\infty 
dM^2\frac{(M^2)^{d_U-2}}{(M^2+\zeta v^2 + \xi \sigma^2)^2}\ .
\label{sigma}
\ee
or, performing the integral explicitly,
\be
\sigma^2 = \frac{1}{4} \Gamma(d_U-1)\Gamma(3-d_U)(\mu_U^2)^{2-d_U} v^4 
(\zeta v^2 + \xi \sigma^2)^{d_U-3} \ ,
\label{sigmai}
\ee
which can be solved for $\sigma^2$ (numerically if $\zeta\neq 0$ or 
analytically if $\zeta=0$).
 
The induced mass gap in the unparticle continuum is now 
\be
m_g^2 = \zeta v^2 + \xi \sigma^2\ ,	
\ee
and it is clear that this mass gap will cutoff the IR divergence of
${\cal O}_U$ even for $\zeta=0$, solving therefore the infrared
problem.  Note that $\sigma\neq 0$ only if $v\neq 0$ so that the mass
gap is in any case associated with EWSB.

\section{Unparticle Plasmon Excitation}

We begin by writing down explicitly the infinite mass matrix that
mixes the (real) neutral component $h^0$ of the Higgs with the
deconstructed tower of unparticle scalars, $\varphi_n$.  The different
matrix elements are:
\bea
\label{massmatrixhh}
M_{hh}^2 & = & 2\lambda v^2 \equiv m_{h0}^2\ ,\\
\label{massmatrixhn}
M_{h n}^2 & = & \kappa_U v F_n\frac{M_n^2 + \xi 
\sigma^2}{M_n^2+m_g^2}\equiv A_n\ ,\\
M_{nm}^2 & = &  (M_n^2+m_g^2)\delta_{nm}+\frac{1}{2}\kappa_U^2 \xi 
v^4\frac{F_nF_m}{(M_n^2+m_g^2)(M_m^2+m_g^2)}\nonumber\\
&&\equiv (M_n^2+m_g^2)\delta_{nm}+a_na_m\ .
\label{massmatrixnm}
\eea

It is a simple matter to obtain the $hh$-entry of the inverse
(infinite matrix) propagator associated to this infinite mass
matrix. Already taking its continuum limit we obtain:
\be
\label{invprop}
iP_{hh}(p^2)^{-1}= p^2 - m_{h0}^2 + J_2(p^2)-\frac{1}{2}\xi v^2
\frac{[J_1(p^2)]^2}{1+\frac{1}{2}\xi v^2 J_0(p^2)}\ ,
\ee
where we have used the integrals
\bea
J_k(p^2)&\equiv&\int_0^\infty G_U(M^2,p^2) (M^2+\xi\sigma^2)^{k} dM^2
\nonumber\\
&=&\frac{v^2}{p^4}\left(\frac{\mu_U^2}{m_g^2}\right)^{2-d_U}\Gamma(d_U-1)
\Gamma(2-d_U)\left\{\left(1-\frac{p^2}{m_g^2}\right)^{d_U-2}
(p^2-m_g^2+\xi\sigma^2)^k\right.\nonumber\\
&-&\left.
\left[1+(2-d_U)\frac{p^2}{m_g^2}\right](\xi\sigma^2-m_g^2)^k-k \ 
p^2(\xi\sigma^2-m_g^2)^{k-1}\right\}\ ,
\label{Jk}
\eea
with integer $k$ and where $G_U(M^2,p^2)$ is:
\be
G_U(M^2,p^2)\equiv\frac{v^2(\mu_U^2/M^2)^{2-d_U}}{(M^2+m_g^2-p^2)(M^2+m_g^2)^2}
\ .
\ee
These integrals are real for $p^2<m_g^2$ but they develop an 
imaginary part for $p^2>m_g^2$. This imaginary part will be important 
later on when we discuss the spectral function associated to 
$P_{hh}(p^2)$.
The final expression for the inverse propagator with all the integrals
explicitly performed is lengthy and not very illuminating. Although
the integrals in (\ref{invprop}) diverge for $p^2\rightarrow m_g^2$,
the combination entering (\ref{invprop}) is finite.

In contrast with the scenario analyzed in Ref.~\cite{DEQ}, in which
the (real part of the) Higgs-unparticle propagator had a pole associated 
with a Higgs
(with non-standard couplings), the propagator (\ref{invprop}) has an
additional pole associated with the unparticle continuum.  In order to
understand the origin of this additional pole consider the unparticle
submatrix (\ref{massmatrixnm}). It has a simple form (a diagonal part
plus a rank-1 correction) that allows one to find a particularly
interesting eigenvalue $\omega_{p0}^2$ (and eigenvector $\{r_n\}$)
that satisfy
\be
1+\sum_n\frac{a_n^2}{M_n^2+m_g^2-\omega_{p0}^2}=0\ ,
\label{plasmon}
\ee
and
\be
r_n=\frac{a_n}{N_p(\omega_{p0}^2-M_n^2-m_g^2)}\ ,
\ee
where $N_p$ is a normalization constant that ensures
$\sum_{n=1}^\infty r_n^2=1$. For sufficiently large values of the
$a_n$'s Eq.~(\ref{plasmon}) has a solution, with $\omega_{p0}^2>m_g^2$
necessarily. Note that this pole can exist due to the presence of the
new quartic coupling $\xi$ and only after EWSB, which gives $a_n\neq
0$. The appearance of this state out of the unparticle continuum is
reminiscent of the appearance of plasmon excitations in condensed
matter physics. In fact, the structure of the unparticle submatrix is
similar to the Hamiltonian that describes the residual long-range
Coulomb interactions induced in a plasma by a probe electromagnetic
wave. Such structure lies at the root of different collective
phenomena in different fields of physics~\cite{fano}.

The previous discussion can be carried over to the continuum limit, in 
which the condition (\ref{plasmon}) takes the form
\be
1+\frac{1}{2}\xi v^2\ {\mathrm P.V.}\left[\int_0^\infty 
G_U(M^2,\omega_{p0}^2)dM^2\right]=
1+\frac{1}{2}\xi v^2\ {\mathrm P.V.}[J_0(\omega_{p0}^2)]
=0\ ,
\label{plasmonc}
\ee
and will be modified only quantitatively by the mixing of unparticles
with the Higgs in the full matrix
(\ref{massmatrixhh})-(\ref{massmatrixnm}).  In general we will expect
two poles, one Higgs-like at $m_h^2$ coming from the unmixed
$m_{h0}^2$, and one plasmon-like at $\omega_p^2$ coming from the
unmixed $\omega_{p0}^2$, both of them somewhat displaced by the
mixing.

\section{Spectral Function Analysis}

In order to study in more detail this interplay between the Higgs and
the unparticle sector it is instructive to examine the spectral
representation of the mixed propagator (\ref{invprop}), which is given by
\be
\rho_{hh}(s) =-\frac{1}{\pi} Im[P_{hh}(s+i\epsilon)]\ ,
\ee
where the limit $\epsilon\rightarrow 0$ is understood. We can easily
calculate this spectral function by using $1/(x+i\epsilon)\rightarrow
{\mathrm P.V.}[1/x] -i\pi\delta(x)$ directly in the integrals $J_k$ of
(\ref{Jk}) to obtain, for $s>m_g^2$,
\be
J_k(s+i\epsilon)=R_k(s)+i I_k(s)\ ,
\ee
with
\bea
R_k(s)&=& \frac{v^2}{s^2}\left(\frac{\mu_U^2}{m_g^2}\right)^{2-d_U} 
\Gamma(d_U-1)\Gamma(2-d_U)\left\{\left(\frac{s}{m_g^2}-1\right)^{d_U-2}
(s-m_g^2+\xi\sigma^2)^k
\cos(d_U\pi) 
\right.\nonumber\\
&-&\left.
\left[1+(2-d_U)\frac{s}{m_g^2}\right](\xi\sigma^2-m_g^2)^k-k \ 
s\ (\xi\sigma^2-m_g^2)^{k-1}\right\}\ , \nonumber\\
I_k(s)&=&\pi 
\frac{v^2}{s^2}(s+\xi\sigma^2-m_g^2)^k
\left(\frac{\mu_U^2}{s-m_g^2}\right)^{2-d_U}\ .
\eea

As in the case of Ref.~\cite{DEQ} there are two qualitatively different
cases, depending on whether the Higgs mass $m_h$ is larger or
smaller than $m_g$. For $m_h<m_g$, the spectral function is explicitly
given by
\be
\label{rho}
\rho_{hh}(s)= \frac{1}{K^2(m_h^2)} \delta(s-m_h^2)+\theta(s-m_g^2) 
\frac{T_U(s)}{\mathcal{D}^2(s)+\pi^2T_U^2(s)}
\ ,
\ee
where $\mathcal{D}(s)$ and $T_U(s)$ are the real and imaginary parts of 
$iP_{hh}(s+i \epsilon)^{-1}$ when $s>m_g^2$:
\be
iP_{hh}(s+i \epsilon)^{-1}=
\mathcal{D}(s)+i\ T_U(s)\ .
\ee
More explicitly, one finds
\bea
\label{DS}
\mathcal{D}(s)&=&s-m_{h0}^2 + R_2(s)\\
&&\nonumber\\
&-&\frac{1}{2N(s)}\xi 
v^2\left\{\left[1+\frac{1}{2}\xi v^2 
R_0(s)\right]\left[R_1(s)^2-I_1(s)^2\right]+\xi 
v^2I_0(s)R_1(s)I_1(s)\right\}\ ,\nonumber\\
&&\nonumber\\
T_U(s)&= & \frac{v^2}{s^2N(s)}  
(s-\xi \sigma^2-m_g^2)^2
\left(\frac{\mu_U^2}{s-m_g^2}\right)^{2-d_U}
\ ,\label{TU}
\eea
with
\be
N(s)\equiv \left[1+\frac{1}{2}\xi v^2
R_0(s)\right]^2+\left[\frac{1}{2}\xi v^2
I_0(s)\right]^2\ .
\ee
Finally
\be
\label{Ks0}
K^2(s_0)\equiv 
\left.\frac{d}{ds}\mathcal{D}(s)\right|_{s=s_0}\ .
\ee
An explicit expression for $K^2(s_0)$ can be obtained directly from
$\mathcal{D}(s)$ above, but we do not reproduce it here.

One can check that the spectral function (\ref{rho}) is properly
normalized:
\be
\int_0^\infty \rho_{hh}(s) ds =1\ .
\label{norm}
\ee

\begin{figure}[tb]
\includegraphics[width=11.cm,height=15.cm,angle=-90]{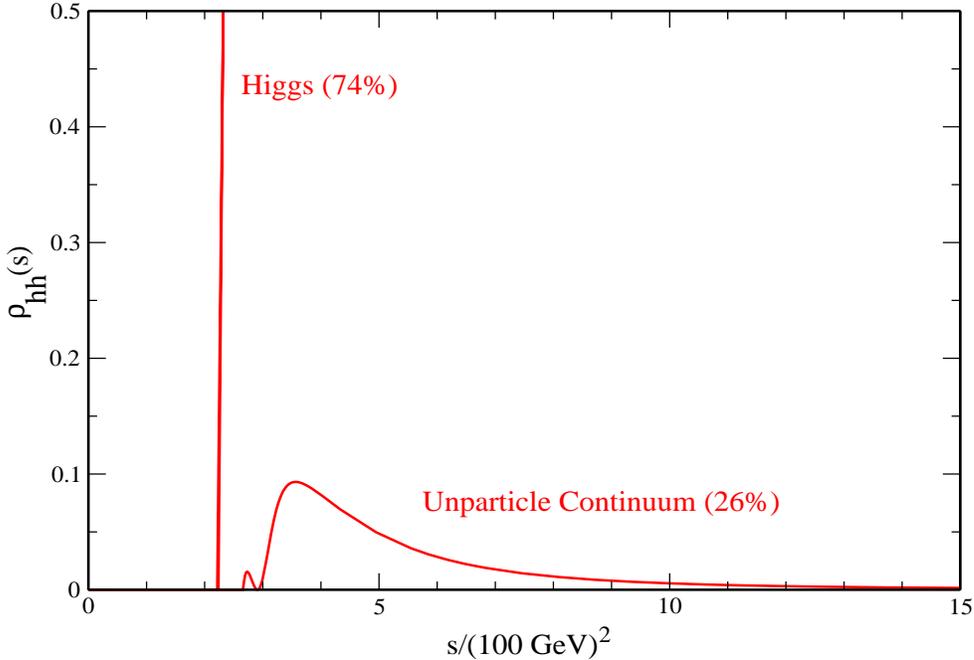}
\caption{\label{rho0} Spectral function with a Higgs below $m_g$,
obtained for the case $\zeta=0.4$, $\xi=0.1$, $m^2=0$,
$\kappa_U=v^{2-d_U}$ and $d_U=1.2$.  The percentage of Higgs
composition of the isolated pole and of the unparticle continuum is
given in parenthesis.}
\end{figure}

The physical interpretation of this spectral function is the standard one:
Let us call $|h\rangle$ the Higgs interaction eigenstate and 
$|u,M\rangle$ the unparticle interaction eigenstates (a continuous 
function of $M$) and $|H\rangle$, $|U,M\rangle$ the respective 
mass eigenstates after EWSB. Then one has
\bea
|\langle H|h \rangle|^2 &=& \frac{1}{K^2(m_h^2)}\ ,\nonumber\\
|\langle U,M |h \rangle|^2 &=&
\frac{T_U(M^2)}{\mathcal{D}^2(M^2)+\pi^2T_U^2(M^2)}\ ,
\eea
so that $\rho_{hh}$ describes in fact the Higgs composition of the
isolated pole and the unparticle continuum. The proper normalization
(\ref{norm}) is simply a consequence of the proper normalization of
$|h \rangle$, {\em i.e.} $|\langle h|h \rangle|^2=1$.  From the simple
form of $T_U(s)$ in (\ref{TU}) we can see directly that for
$M_0^2=m_g^2+\xi \sigma^2$ the spectral function is zero,
corresponding to an unparticle state $|U,M_0\rangle$ which has
$\langle h|U,M_0\rangle=0$. The amount of $|h\rangle$ admixture in any
state is important because it will determine key properties of that
state, like its coupling to gauge bosons, that are crucial for its
production and decay.

Fig.~\ref{rho0} shows the spectral function for a case with $m_h<m_g$.
The parameters have been chosen as follows: $d_U=1.2$,
$\kappa_U=v^{2-d_U}$, $m^2=0$, $\zeta=0.4$ and $\xi=0.1$. We see a
Dirac delta at $m_h^2=(152\ {\rm GeV})^2$, a mass gap for the
unparticle continuum at $m_g^2=(163\ {\rm GeV})^2$, and a zero at
$M_0^2=(171\ {\rm GeV})^2$. There is also a plasmon-like resonance at
$\omega_p^2=(176\ {\rm GeV})^2$ but it is not very conspicuous in this
particular case.  In parenthesis we give the percentage of Higgs
composition in the isolated resonance and in the continuum: it is
simply given by the integral of $\rho_{hh}(s)$ in the corresponding
region. We see that the Higgs has lost some of its original Higgs
composition due to mixing with the unparticles (as in the usual
singlet dilution) while the unparticle continuum gets the lost Higgs
composition spread above $m_g$ (in a continuum way reminiscent of the
models considered in~\cite{EG}).

\begin{figure}[tb]
\includegraphics[width=11.cm,height=15.cm,angle=-90]{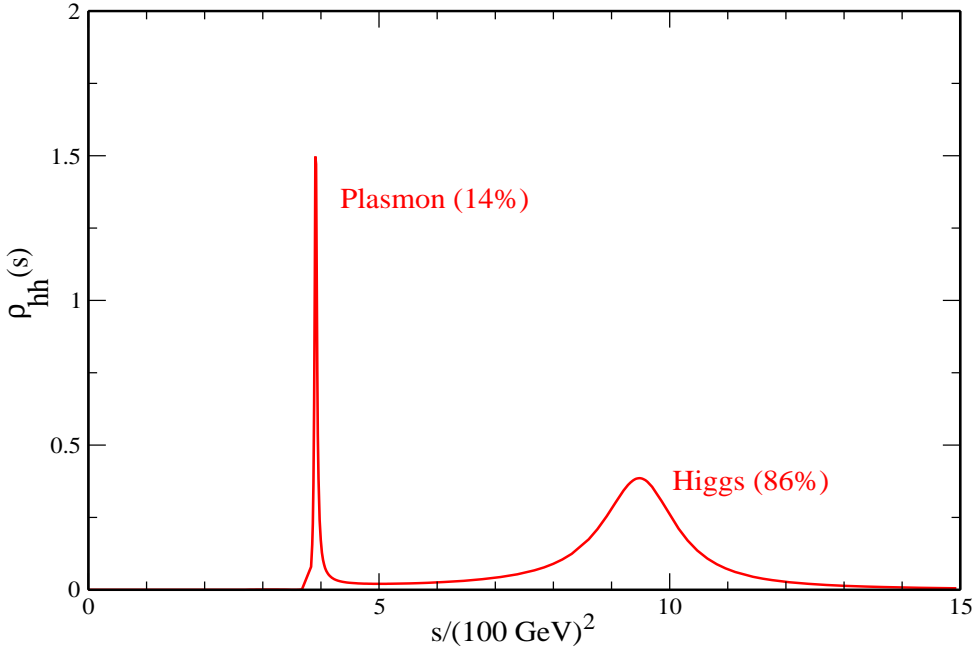}
\caption{\label{rho1} Spectral function with plasmon and Higgs above
$m_g$, obtained for the case $\zeta=0.3$, $\xi=0.2$, $m^2=-1.5 (100\
{\rm GeV})^2$, $\kappa_U=v^{2-d_U}$ and $d_U=1.1$. The percentage of
Higgs composition of each resonance is given in parenthesis.}
\end{figure}

The plasmon-like resonance can be seen much more clearly in other
cases, like the one shown in Fig.~\ref{rho1}, which has $m_h>m_g$. It
corresponds to $d_U=1.1$, $\kappa_U=v^{2-d_U}$, $m^2=-1.5 (100\
{\mathrm GeV})^2$, $\zeta=0.3$ and $\xi=0.2$ and has a mass gap at
$m_g^2=(164\ {\rm GeV})^2$, a Higgs resonance at $m_h^2=(307\ {\rm
GeV})^2$ and a plasmon-like spike at $\omega_p^2=(198\ {\rm
GeV})^2$. There is also a zero at $M_0^2=(188\ {\rm GeV})^2$ right
below the plasmon resonance, but it cannot be discerned in the plot
due to the scale of the figure. We give again in parenthesis the Higgs
composition of the Higgs and plasmon resonances.  For $m_h>m_g$, the
spectral function is given by the second part of (\ref{rho}) only,
without a Dirac delta-function, and there is no separate $|H\rangle$
state.

The shape of the continuum around the resonances at
$s_r=\{m_h^2,\omega_p^2\}$ can be obtained directly from the spectral
density (\ref{rho}) by writing
\be 
\label{nearpl}
\mathcal{D}(s)\simeq (s-s_r)K^2(s_r)\ , 
\ee 
where $K^2(s_r)$ is defined in Eq.~(\ref{Ks0}).  In this case, with
$s_r>m_g^2$, one should be careful about using the principal value
definition of the integrals entering $\mathcal{D}(s)$ to properly
calculate its derivative at $s_r$. Substituting (\ref{nearpl}) in the
spectral function (\ref{rho}), we see that the resonances have a
Breit-Wigner shape of width $\Gamma_r$ given by
\be 
\label{widthpl} 
\frac{\Gamma_r}{\sqrt{s_r}} = \frac{\pi T_{U}(s_r)}{s_r K^2(s_r)}\ .  
\ee 

\section{Conclusions}

An unparticle sector could be explored experimentally in a very
interesting way if it is coupled to the Standard Model directly
through the Higgs $|H|^2$ operator. In this paper we have revisited
such couplings of the Higgs to an unparticle scalar operator ${\cal
O}_U$ of non-integer dimension $d_U$. We have expanded upon our
previous work~\cite{DEQ} by considering a new way of solving the
infrared problem that affects the expectation value of ${\cal O}_U$
for $d_U<2$~\cite{DEQ} that is generated by EWSB. We have shown how a
scale-invariant unparticle self-coupling~\footnote{The importance of
unparticle self-interactions for phenomenology has been emphasized 
in~\cite{Strassler}.} can in fact generate a mass gap $m_g$ for
unparticles that acts as an IR cutoff to give a finite $\langle{\cal
O}_U\rangle$.

In addition to solving the IR problem, the new coupling can induce
after EWSB a new resonance in the unparticle continuum through a
mechanism quite similar to those giving rising to plasmon resonances
in condensed matter systems~\cite{fano}. The mass mixing of
unparticles with the Higgs after EWSB results in a spectrum of states
with some admixture of Higgs that will dictate some of their
production and decay properties. One can distinguish two generic types
of spectra. In the first, there is an isolated state below the mass gap, 
which one would typically identify with the Higgs boson although
it will carry some unparticle admixture that will change its
properties with respect to a SM Higgs (e.g. the coupling to gauge
bosons will be reduced). Beyond the mass gap there will be an
unparticle continuum (possibly with a large plasmon resonance) that
will be accessible experimentally through its Higgs admixture.

In the second type of spectrum, the Higgs mass will be above the mass gap
and the Higgs resonance will in fact merge with the unparticle
continuum acquiring a significant width. In addition to this resonance
a large plasmon resonance can also be present. Both resonances will
have some Higgs admixture so that both could show up experimentally as
Higgses with non-standard properties.

\subsection*{Acknowledgments}

\noindent 
Work supported in part by the European Commission under the European
Union through the Marie Curie Research and Training Networks ``Quest
for Unification" (MRTN-CT-2004-503369) and ``UniverseNet"
(MRTN-CT-2006-035863); by the Spanish Consolider-Ingenio 2010
Programme CPAN (CSD2007-00042); by a Comunidad de Madrid project
(P-ESP-00346) and by CICYT, Spain, under contracts FPA 2007-60252 and
FPA 2005-02211.

\end{document}